\documentclass[twocolumn, prd, showpacs,
floats]{revtex4}
\usepackage{graphicx}
\usepackage{dcolumn}
\usepackage{bm}
\usepackage{amsmath}

\begin{document}

\title{Torsion in Gauge Theory}

\author{H.T. Nieh}
\email{nieh@tsinghua.edu.cn}
\affiliation{Institute for Advanced Study, Tsinghua University, Beijing 100084, China}

\begin{abstract}

The potential conflict between torsion and gauge symmetry in the Riemann-Cartan curved spacetime was noted by Kibble in his 1961 pioneering paper, and has since been discussed by many authors. Kibble suggested that, to preserve gauge symmetry, one should forgo the covariant derivative in favor of the ordinary derivative in the definition of the field strength $F_{\mu\nu}$ for massless gauge theories, while for massive vector fields covariant derivatives should be adopted. This view was further emphasized by Hehl and collaborators in their influential 1976 review paper. We address the question of whether this deviation from \textit{normal} procedure by forgoing covariant derivatives in curved spacetime with torsion could give rise to inconsistencies in the theory, such as the quantum renormalizability of a realistic interacting theory. We demonstrate in this note the one-loop renormalizability of a realistic gauge theory of gauge bosons interacting with Dirac spinors, such as the SU(3) chromodynamics, for the case of a curved Riemann-Cartan spacetime with totally antisymmetric torsion. This affirmative confirmation is one step towards providing justification for the assertion that the flat-space definition of the gauge field strength should be adopted as the proper definition.

\end{abstract}


\date{December 4, 2017}
\maketitle

\vspace{3mm}

\section{Introduction}
In the formulation of a physical theory in curved spacetime, the \textit{normal} procedure is to replace the ordinary derivative with the corresponding covariant derivative. For a gauge theory in the Riemannian spacetime, because the connection is symmetric, the normal procedure yields a field strength tensor $F_{\mu\nu}$ in the form of its flat-space expression, which is gauge symmetric. But in a Riemann-Cartan spacetime with torsion, the connection being non-symmetric, this same procedure gives rise to an additional torsion term in the gauge field strength tensor that violates gauge symmetry. Torsion naturally appears in the Einstein-Cartan-Kibble-Sciama theory of gravitation \cite{Kibble, Sciama}. The potential conflict of torsion with gauge symmetry was already noticed by Kibble \cite{Kibble} in his original paper, and has since been discussed by a number of authors \cite{Hehl, Hojman, Ni, Benn, Sabbata, Soleng, Hammond, Puntigam, Andrade, Itin} with various alternatives. Kibble \cite{Kibble} himself took the view that, to preserve gauge symmetry, one should forgo the covariant derivative in favor of the ordinary derivative in the definition of the field strength $F_{\mu\nu}$ for massless gauge theories, while for massive vector fields covariant derivatives should be adopted. This view was adopted by Hehl and collaborators \cite{Hehl} in their influential 1976 review paper. Since all other alternatives suggested by various authors did not seem to hold up, Kibble's original view has been tacitly accepted without further deliberations, seemingly as consensus by default. The situation is the following. We are facing two alternative choices of $F_{\mu\nu}$, one with the torsion term and the other without. It is uncertain whether the latter alternative, forcing gauge symmetry by deviating from the \textit{normal} procedure of defining $F_{\mu\nu}$ through covariant derivatives, would cause inconsistency or non-covariant issues in a realistic \textit{quantum} gauge theory, such as the SU(3) quantum chromodynamics, in a curved Riemann-Cartan spacetime, in which all other operations, such as gauge fixing and the ensuing ghost supplementation, all follow \textit{normal} covariant procedures. This uncertainty, at least, needs a clarification. We report in this paper our findings regarding system consistency for the two alternative $F_{\mu\nu}$ cases within the framework of the Kibble-Sciama scheme as well as the renormalizability question. We will first show that the system of field equations, even at the classical level, is inconsistent if the field strength $F_{\mu\nu}$ takes the gauge non-symmetric form, while it is consistent with the gauge symmetric $F_{\mu\nu}$. This clearly rules out the gauge non-symmetric version of $F_{\mu\nu}$. We will next demonstrate, using the gauge-invariant background-field method \cite{DeWitt, Honerkamp, 't Hooft, Abbott}, in conjunction with the heat-kernel technique and dimensional regularization, that the theory is renormalizable at the one-loop level in the case of the gauge-symmetric field strength $F_{\mu\nu}$, in a Riemann-Cartan spacetime with totally antisymmetric torsion. These findings provide substantiation for the choice of the gauge symmetric version of $F_{\mu\nu}$ and validates the view of Kibble \cite{Kibble} and Hehl et al.\cite{Hehl}.

\section{Sciama-Kibble Scheme}

The genesis of the Kibble-Sciama \cite{Kibble,Sciama} theory can be traced back to the formulation of the Dirac equation in curved spacetime by Weyl \cite{Weyl} and Fock \cite{Fock}. The vierbein fields $e^{a}_{~\mu}$ were introduced by Weyl and Fock to provide local coordinate basis for defining the Dirac spinor, and the spin connection field $\omega^{ab}_{~~\mu}$ as the gauge potential for the SO(3,1) group of local Lorentz transformations of the Dirac spinor. Utiyama \cite{Utiyama} demonstrated that Einstein's Riemannian theory of gravitation can be regarded as a gauge theory of the SO(3,1) Lorentz group when the corresponding gauge potential, the spin connection $\omega^{ab}_{~~\mu}$, is identified with the Ricci coefficients of rotation \cite{Fock} in terms of the vierbein fields $e^{a}_{~\mu}$. Sciama and Kibble \cite{Kibble,Sciama} took the step of treating the spin connection field $\omega^{ab}_{~~\mu}$, in the spirit of a genuine Lorentz group gauge theory, as independent dynamic variable to be determined by the theory, instead of being identified with the Ricci coefficients. The coupling of the spin connection to the Dirac spinors, for example, gives rise to torsion.

The metric tensor $g_{\mu\nu}$ in the Kibble-Sciama scheme is defined by
$$g_{\mu\nu}=\eta_{ab}e_{~\mu}^{a}e_{~\nu}^{b}, \eqno(1) $$
\noindent where $\eta_{ab}=(1,-1,-1,-1)$, and the covariant derivatives with respect to both local Lorentz transformations and general coordinate transformations, for generic $\chi_{a}^{~\lambda}$ and $\chi_{~\nu}^{a}$, are defined according to
$$\nabla_{\mu}\chi_{a}^{~\lambda}=\chi_{a~,\mu}^{~\lambda}-\omega_{~a\mu}^{b}\chi_{b}^{~\lambda}+\Gamma_{~\nu\mu}^{\lambda}\chi_{a}^{~\nu}, \eqno(2) $$
$$\nabla_{\mu}\chi_{~\nu}^{a}=\chi_{~\nu,\mu}^{a}+\omega_{~b\mu}^{a}\chi_{~\nu}^{b}-\Gamma_{~\nu\mu}^{\lambda}\chi_{~\lambda}^{a}. \eqno(3) $$
\noindent Kibble \cite{Kibble} chose the affine connection
$$\Gamma_{~\mu\nu}^{\lambda}=e_{a}^{~\lambda}(e_{~\mu,\nu}^{a}+\omega_{~b\nu}^{a}e_{~\mu}^{b}), \eqno(4) $$
\noindent so that it is metric compatible, meaning
$$\nabla_{\lambda}e_{~\mu}^{a}=0, \eqno (5) $$
$$\nabla_{\lambda}e_{a}^{~\mu}=0, \eqno (6) $$
\noindent and, consequently,
$$\nabla_{\lambda}g^{\mu\nu}=0, \eqno (7) $$
$$\nabla_{\lambda}g_{\mu\nu}=0. \eqno (8) $$

In the presence of torsion, which is defined as
$$C_{~\mu\nu}^{\lambda}=\Gamma_{~\mu\nu}^{\lambda}-\Gamma_{~\nu\mu}^{\lambda}, \eqno (9) $$
\noindent the metric compatibility relations (7) and (8) imply that the connection is of the general form:
$$\Gamma_{~\mu\nu}^{\lambda}=\frac{1}{2}g^{\lambda\rho}(g_{\rho\mu,\nu}+g_{\nu\rho,\mu}-g_{\mu\nu,\rho})+Y_{~\mu\nu}^{\lambda}, \eqno (10) $$
\noindent where the contortion tensor $Y_{~\mu\nu}^{\lambda}$ is given by
$$Y_{~\mu\nu}^{\lambda}=\frac{1}{2}(C_{~\mu\nu}^{\lambda}+C_{\mu\nu}^{~~\lambda}+C_{\nu\mu}^{~~\lambda}). \eqno (11) $$

\section{System of Gluons interacting with Quarks in Sciama-Kibble Scheme}

 For notational convenience of presentation, we shall consider the specific case of the SU(3) chromodynamics, in which the gauge gluons interacting with a triplet of massless spinor quarks. Let the gauge field be denoted by $A_{\mu}^{\underline{a}}$, where the index \underline{a} runs from 1 to 8. It is convenient to adopt the group algebraic notation
$$A_{\mu}=A_{\mu}^{\underline{a}}T^{\underline{a}}, \eqno (12) $$

\noindent where $T^{\underline{a}}$, for concreteness, are the familiar 3x3 $\frac{1}{2}\lambda_{\underline{a}}$ Gell-Mann matrices satisfying the algebra, with the totally antisymmetric $f^{\underline{abc}}$ being the SU(3) group structure constants,
$$[T^{\underline{a}},T^{\underline{b}}]=if^{\underline{abc}}T^{\underline{c}}. \eqno (13) $$
In flat space, the field strength is given by
$$F_{\mu\nu}=\partial_{\mu}A_{\nu}-\partial_{\nu}A_{\mu}-ig[A_{\mu},A_{\nu}].  \eqno (14)$$
\noindent In curved space, the natural definition for the field strength is to follow the normal procedure of replacing the partial derivative by the appropriate covariant derivative, like
$$\partial_{\mu}A_{\nu}\rightarrow\partial_{\mu}A_{\nu}-\Gamma_{~\nu\mu}^{\lambda}A_{\lambda}. \eqno (15) $$
\noindent In the Riemannian space, the connection being symmetric, the connection terms cancel when the replacement (14) is made in (13), leaving the expression for $F_{\mu\nu}$ unchanged. In a Riemann-Cartan space, the connection is non-symmetric, and the field strength $F_{\mu\nu}$ resulted from the replacement is of the form
$$F_{\mu\nu}=\partial_{\mu}A_{\nu}-\partial_{\nu}A_{\mu}+C_{~\mu\nu}^{\lambda}A_{\lambda}-ig[A_{\mu},A_{\nu}]. \eqno (16) $$
\noindent The additional torsion term in (15) violates gauge invariance. To preserve gauge symmetry, an alternative is to forgo the torsion term in (15) and adopt the flat-space expression (14) as the definition of the field strength $F_{\mu\nu}$.
We now consider the system of SU(3) gauge bosons interacting with spinor quarks in the background of the curved Riemann-Cartan space as described above in the Kibble-Sciama scheme. We first check the consistency of the system of field equations with the two alternative versions of the field strength $F_{\mu\nu}$, (14) and (16). For convenience, we shall consider a massless spinor quark, which is denoted by $\psi$. The action for the system is of the form, with trace over color index understood,
$$\begin{array}{rl}W=\displaystyle\int d^{4}xh[-\frac{1}{4}F^{\mu\nu}F_{\mu\nu}\\
+\frac{1}{2}(\bar{\psi}i\gamma^{a}e_{a}^{~\mu}D_{\mu}\psi-\bar{\psi}\bar{D}_{\mu}i\gamma^{a}e_{a}^{~\mu}\psi)], \end{array} \eqno (17) $$
\noindent where $h=\det{e}_{~\mu}^{a}$, and
$$D_{\mu}=\partial_{\mu}-\frac{i}{4}\sigma_{ab}\omega_{~~\mu}^{ab}+igA_{\mu}, \eqno (18)  $$
$$\bar{D}_{\mu}=\overleftarrow{\partial}_{\mu}-igA_{\mu}^{\underline{a}}T^{\underline{a}}+\frac{i}{4}\sigma_{ab}\omega_{~~\mu}^{ab}, \eqno (19) $$
\noindent with $\sigma_{ab}=\frac{i}{2}[\gamma_{a},\gamma_{b}]$ \cite{Bjorken}. We note that $F^{\mu\nu}=g^{\mu\lambda}g^{\nu\rho}F_{\lambda\rho}$.
\noindent For proper normalization of the $F^{\mu\nu}F_{\mu\nu}$ term in the Lagrangian, there should be a factor of $\frac{1}{C_{2}(R)}$, which is defined by
$$tr(T^{\underline{a}}T^{\underline{b}})=C_{2}(R)\delta^{\underline{ab}}.  \eqno (20) $$
\noindent For convenience, we have omitted this normalization factor, but it will be taken into account when we consider renormalization counter terms.
The Lagrangian in the action (17) is invariant under local Lorentz transformations, general coordinate transformations as well as local scale transformations, the latter being defined, with the proper scale weights for the various fields, by
$$e_{a}^{~\mu}\rightarrow e^{-\Lambda(x)}e_{a}^{~\mu},$$
$$e_{~\mu}^{a}\rightarrow e^{\Lambda(x)}e_{~\mu}^{a},$$
$$\psi(x)\rightarrow e^{-\frac{3}{2}\Lambda(x)}\psi(x),$$
$$A_{\mu}(x)\rightarrow A_{\mu},$$
$$\omega_{~~\mu}^{ab}(x)\rightarrow \omega_{~~\mu}^{ab}(x).$$
\noindent We note the scale invariance of the Dirac Lagrangian in (17) without explicit appearance of a Weyl scale gauge field; even if such a gauge field were introduced in the covariant derivative $D_{\mu}$, it would drop out from the Lagrangian, due to cancellation between the two hermitian conjugate terms, and would not appear in the ensuing field equation for the Dirac field $\psi(x)$.
\noindent Regarding the Maxwell field strength $F_{\mu\nu}$, we consider separately its two alternative versions, namely (14) and (16), respectively.

\section{Case (I) Gauge Non-symmetric $F_{\mu\nu}$.}

First, we consider the version with the field strength $F_{\mu\nu}$ containing the torsion term, namely,
$$F_{\mu\nu}=\partial_{\mu}A_{\nu}-\partial_{\nu}A_{\mu}+C_{~\mu\nu}^{\lambda}A_{\lambda}-ig[A_{\mu},A_{\nu}], \eqno (16) $$
\noindent which is not gauge invariant. The Euler-Lagrange equation for the Dirac field can be obtained straightforwardly from (17). On account of the relation
$$h^{-1}\partial_{\mu}h=\Gamma_{~\lambda\mu}^{\lambda}=\Gamma_{~\mu\lambda}^{\lambda}+C_{~\lambda\mu}^{\lambda}, \eqno (21)  $$
\noindent and the commutation properties of the Dirac gamma matrices \cite{Bjorken}, we obtain \cite{Nieh,Obukhov,Lasenby} the field equation for the Dirac field $\psi$,
$$i\gamma^{a}e_{a}^{\mu}(D_{\mu}+\frac{1}{2}C_{~\lambda\mu}^{\lambda})\psi=0, \eqno (22)  $$
\noindent where $D_{\mu}$ is given in (19). We know that the Lagrangian in the action (17) is scale invariant. The Dirac equation (22) is thus expected to be scale invariant. We have, by its construction according to (8), the connection $\Gamma_{~\mu\nu}^{\lambda}$ has the following scale transformation property
$$\Gamma_{~\mu\nu}^{\lambda}\rightarrow \Gamma_{~\mu\nu}^{\lambda}+\delta_{~\mu}^{\lambda}\Lambda_{,\nu}, \eqno (23)  $$
\noindent which implies
$$C_{~\lambda\mu}^{\lambda}\rightarrow C_{~\lambda\mu}^{\lambda}+3\Lambda_{,\mu}. \eqno (24)  $$
\noindent We denote
$$B_{\mu}=\frac{1}{3}C_{~\lambda\mu}^{\lambda}. \eqno (25)  $$
\noindent It transforms as an effective \textit{Weyl gauge field} for local scale transformations \cite{Nieh,Obukhov,Lasenby}
$$B_{\mu}\rightarrow B_{\mu}+\Lambda_{,\mu}. \eqno (26)  $$
\noindent The Dirac equation (22) is then expressed as
$$i\gamma^{a}e_{a}^{\mu}(D_{\mu}+\frac{3}{2}B_{\mu})\psi=0. \eqno (27)  $$
\noindent So, indeed, the massless Dirac equation written in this form shows explicit scale invariance, and with the proper scale weight $\frac{3}{2}$ for the Dirac field $\psi$.
The Euler-Lagrange equation for the gauge field is obtained straightforwardly. It is of the form
$$(\nabla_{\nu}+3B_{\nu})F^{\mu\nu}=gJ^{\mu}, \eqno (28)  $$
\noindent where the covariant derivative $\nabla_{\nu}$ is defined as in
$$\nabla_{\nu}F^{\mu\nu}=\partial_{\nu}F^{\mu\nu}+\Gamma_{~\lambda\nu}^{\mu}F^{\lambda\nu}+\Gamma_{~\lambda\nu}^{\nu}F^{\mu\lambda}-ig[A_{\nu},F^{\mu\nu}], \eqno (29) $$
\noindent and the current $J_{\mu}$ given by
$$J^{\mu}=\bar{\psi}\gamma^{a}e_{a}^{~\mu}T^{\underline{a}}\psi T^{\underline{a}}. \eqno (30)  $$
In the presence of torsion, the field equation (28) is not gauge invariant. We would like to check whether current conservation is valid and whether the system of field equations, namely (27) and (28), are mutually consistent. As a consequence of the Dirac equation (27) and its hermitian conjugate equation for $\bar{\psi}$, it is straightforward to verify that the current $J^{\mu}$ given by (30)is indeed conserved,
$$(\nabla_{\mu}+3B_{\mu})J^{\mu}=0. \eqno (31)  $$
\noindent Consistency of (28) with this current conservation equation (31), which follows directly from the Dirac equation (27), requires that
$$(\nabla_{\mu}+3B_{\mu})(\nabla_{\nu}+3B_{\nu})F^{\mu\nu}=0. \eqno (32)  $$
\noindent Making use of the anti-symmetry of $F^{\mu\nu}$, it is straightforward, though tedious, to show that
$$\begin{array}{rl}(\nabla_{\mu}+3B_{\mu})(\nabla_{\nu}+3B_{\nu})F^{\mu\nu}
=-R_{~\rho\mu\nu}^{\mu}F^{\rho\nu}\\+\frac{1}{2}C_{~\rho\nu}^{\mu}\nabla_{\mu}F^{\rho\nu}+\frac{3}{2}F^{\mu\nu}(\nabla_{\mu}B_{\nu}-\nabla_{\nu}B_{\mu}).\end{array}  \eqno (33)  $$
\noindent For the right-hand side of (33) to vanish, it is necessary, due to its structure, that the second term vanishes. That is, we have to set $C_{~\rho\nu}^{\mu}=0$. This results in $B_{\mu}=o$, and $R^{\mu}_{~\rho\mu\nu}$ being symmetric in $\rho$ and $\nu$ because the connection $\Gamma_{~\mu\nu}^{\lambda}$ now reduces to the Christoffel connection. The three terms on the right-hand side of (29) then all vanish. Consistency of the two field equations of the system (24) and (25) is thus seen to require the vanishing of torsion. The upshot is that the system of field equations is inconsistent for the gauge non-symmetric version (16) of $F_{\mu\nu}$.

\section{Case (II) Gauge Symmetric $F_{\mu\nu}$}

We next consider the case of gauge symmetric $F_{\mu\nu}$
$$F_{\mu\nu}=\partial_{\mu}A_{\nu}-\partial_{\nu}A_{\mu}-ig[A_{\mu},A_{\nu}], $$
\noindent which is the version with the torsion term removed. With this expression for $F_{\mu\nu}$ in the action (17), the field equation for the Dirac field $\psi$ remains the same as (27), resulting in the same current conservation equation (31), while the field equation for the Maxwell field (28) is replaced by
$$(\nabla_{\nu}+3B_{\nu})F^{\mu\nu}-\frac{1}{2}C^{\mu}_{~\rho\nu}F^{\rho\nu}=J^{\mu}. \eqno(34) $$
\noindent Consistency of (34) with (31) requires that $(\nabla_{\mu}+3B_{\mu})$ operating on the left-hand side of (34) vanishes. The result of operating on the first term on the left-hand side of (34) is already found and given by (33). Operating on the second term yields the contribution
$$-\frac{1}{2}C_{~\rho\nu}^{\mu}\nabla_{\mu}F^{\rho\nu}-\frac{1}{2}\nabla_{\mu}C_{~\rho\nu}^{\mu}F^{\rho\nu}-\frac{3}{2}B_{\mu}C^{\mu}_{~\rho\nu}F^{\rho\nu}. \eqno (35)  $$
\noindent Summing the two contributions given in (33) and (35) yields
$$-R_{~\rho\mu\nu}^{\mu}F^{\rho\nu}+\frac{1}{2}[3(B_{\nu,\mu}-B_{\mu,\nu})-\nabla_{\mu}C_{~\rho\nu}^{\mu}]F^{\mu\nu}. \eqno (36) $$
\noindent In the presence of torsion, the antisymmetric part of $R_{~\rho\mu\nu}^{\mu}$ does not vanish, and explixt evaluation gives he result
$$\frac{1}{2}(R_{~\rho\mu\nu}^{\mu}-R_{~\nu\mu\rho}^{\mu})=\frac{1}{2}[3(B_{\nu,\rho}-B_{\rho,\nu})-\nabla_{\mu}C^{\lambda}_{~\rho\nu}]. \eqno (37) $$
\noindent The two contributions from operating $(\nabla_{\mu}+3B_{\mu})$ on the two right-hand side terms of (34) miraculously cancel each other out and the final result is zero. Consistency of the field equations is thus established. The un-pleasing $C^{\mu}_{~\rho\nu}F^{\rho\nu}$ term in the field equation (34) looks formidable, but it actually helped save consistency.
We have thus seen that the system of classical field equations of chromodynamics in the curved Riemann-Cartan space is self consistent when the gauge field strength is defined by the gauge symmetric expression (14), while it is not for the gauge non-symmetric version (16). The latter version is thus ruled out, even at the classical level. We next check whether the gauge symmetric version (13) of the interacting gauge theory, chromodynamics, is one-loop renormalizable.

\section{One-loop Renormalization by Background-field method}

The background-field method \cite{DeWitt,Honerkamp,'t Hooft, Klugberg-Stern, Abbott} is ideally suited to the computation of effective interaction in curved spaces. It has been used to study the renormalization property of gauge theories in curved Riemannian spacetime by various authors \cite{Toms,Omote,Lee,Jack}, establishing renormalizability at one-loop level and beyond. In the case of Riemann-Cartan spacetime, there does not seem to exist investigations in the literature of the renormalizability question of gauge theories. The question in focus is whether torsion could create complications, a question we would like to study. Based on the background-field method, there is the unified super-space computation \cite{Lee} of the one-loop renormalization counter terms, treating both gauge bosons and Dirac fermions within the framework of the Schwinger-DeWitt proper-time representation of the propagator functions \cite{DeWitt,Schwinger}. Rather than using this elegant framework for evaluating the renormalization counter terms, we will instead combine the normal treatment based on heat-kernel technique with 't Hooft's algorithm\cite{'t Hooft, Omote} for extracting one-loop divergences. The one-loop renormalization counter terms arise from four types of loops, the boson gluon loop, the ghost loop, the spinor quark loop, and the mixed gluon-quark loops (quark self energy loop and gluon-quark vertex loops). For the gluon, ghost and quark loops we follow Toms' treatment, which is based on the heat-kernel method (a variant of the Schwinger-DeWitt proper-time method) and dimensional regularization, while for the mixed gluon-quark loops, we make use of the 't Hooft algorithms \cite{'t Hooft,Omote}.
The divergent part of the one-loop effective action is given by an integral of the coefficient $[a_{2}]$ of the heat-kernel expansion \cite{DeWitt,Parker}. Its explicit expression is given by DeWitt \cite{DeWitt} and Gilkey \cite{Gilkey}, in the case of Riemannian space-time. In the case of Riemann-Cartan space-time, the presence of torsion makes the evaluation of the corresponding $[a_{2}]$ quite involved, and there does not seem to exit a definitive result for a general torsion. The special case of totally antisymmetric torsion has been carefully studied by Yajima \cite{Yajima}. It is Yajima's result that we will make use of, and we will thus restrict ourselves to the special case of totally antisymmetric torsion, for which the effective Weyl gauge field vanishes, namely, $B_{\mu}=0$.

Let's denote the classical background fields by $\eta$ and $\hat{A}_{\mu}$, which satisfy the field equations (22) and (28), respectively. We replace in the action (17) the field $A_{\mu}$ by $A_{\mu}+\hat{A}_{\mu}$, and $\psi$ by $\psi+\eta$. In the respective sums, $A_{\mu}$ and $\psi$ (and $\bar{\psi}$) are regarded as quantum fields, while $\hat{A}_{\mu}$ and $\xi$ as classical fields. We remind ourselves that in the action (17) the gauge field strength $F_{\mu\nu}$ is defined by the gauge symmetric expression (14), namely,
$$F_{\mu\nu}=\partial_{\mu}A_{\nu}-\partial_{\nu}A_{\mu}-ig[A_{\mu},A_{\nu}].  $$
\noindent We will also need to add to the action the gauge-fixing term and the corresponding Faddeev-Popov ghost term \cite{Faddeev}. The gauge-fixing term is chosen in accordance to the Landau-DeWitt gauge condition and is given by \cite{Toms}
$$W_{(GF)}=\displaystyle\int d^{4}xh[-\frac{1}{2}(\hat{\nabla}_{\mu}A^{\mu})^{2}], \eqno (38) $$
\noindent where
$$\hat{\nabla}_{\mu}A^{\mu}=\partial_{\mu}A^{\mu}+\Gamma_{~\lambda\mu}^{\mu}A^{\lambda}-ig[\hat{A}_{\mu},A^{\mu}]. \eqno (39) $$
\noindent The gauge-fixing action (38) brakes gauge symmetry if only the quantum field $A^{\mu}$ undergoes gauge transformation, but can be made gauge covariant under suitably combined gauge transformations of both $A^{\mu}$ and $\hat{A}_{\mu}$. The corresponding Faddeev-Popov ghost term can be obtained by changing the integration "variable" in the path integral and is given by
$$W_{(ghost)}=\displaystyle\int d^{4}xh\bar{\zeta}(-\hat{\nabla}_{\mu}\hat{\nabla}^{\mu}-\hat{\nabla}_{\mu}A^{\mu}-A^{\mu}\hat{\nabla}_{\mu})\zeta, \eqno (40) $$
\noindent where the ghost fields $\bar{\zeta}$ and $\zeta$ are Grassmann scalars and carry the same color index as $A_{\mu}$. The gauge-fixing action $W_{(GF)}$ and Faddeev-Popov ghost action $W_{(ghost}$ are to be added to the action $W$, given by (17), to form the total action.
Expand the action in powers of the quantum fields $A_{\mu}$ and $\psi$. The coefficients of terms linear in quantum fields vanish, as a result of the classical field equations (21 and (28). The terms quadratic in the quantum fields (including the ghost fields) give rise to the quark-loop and gluon-loop contributions. They are also sufficient for evaluating the mixed gluon-quark loops in accordance to the 't Hooft's algorithm. The terms quadratic in quantum fields, up to a total divergence term in the integrand, are exhibited in
$$\begin{array}{rl}W^{(2)}=\displaystyle\int d^{4}x h\{(-\frac{1}{4}\tilde{F}^{\mu\nu}\tilde{F}_{\mu\nu}-2iA_{\mu}[\hat{F}^{\mu\nu},A_{\nu}])\\
-\frac{1}{2}({\hat\nabla}_{\mu}A^{\mu})^{2}+{\bar \psi}i\gamma^{\mu}(\hat{D}_{\mu})\psi+{\bar \psi}i\gamma^{\mu}iA_{\mu}\eta+\bar{\eta}i\gamma^{\mu}iA_{\mu}\psi\\
-\frac{1}{2}(\hat{\nabla}_{\mu}A^{\mu})^{2}+{\bar\zeta}(-\hat{\nabla}_{\mu}\hat{\nabla}^{\mu})\zeta\},\end{array} \eqno (41)$$
\noindent where $\hat{\nabla}_{\mu}A^{\mu}$ is given in (39) and
$$\tilde{F}_{\mu\nu}=\partial_{\mu}A_{\nu}-\partial_{\nu}A_{\mu}-ig[\hat{A}_{\mu},A_{\nu}]+ig[\hat{A}_{\nu},A_{\mu}], \eqno (42) $$
$$\gamma^{\mu}=\gamma^{a}e_{a}^{~\mu}, \eqno (43) $$
$$\hat{D}_{\mu}=\partial_{\mu}-\frac{i}{4}\sigma_{ab}\omega_{~~\mu}^{ab}+ig\hat{A}_{\mu}. \eqno (44) $$

\section{Fermion Loop}

The term in $W^{(2)}$ that gives rise to the pure fermion loop is quadratic in the quantum fermion fields, namely,
$$\displaystyle\int d^{4}xh\bar{\psi}i\gamma^{\mu}\hat{D}_{\mu}\psi. $$
\noindent The effective action due to the fermion loop is given by \cite{Toms,Parker}
$$\Gamma_{fermion-loop}=-iln Det(i\gamma^{\mu}\hat{D}_{\mu}). \eqno (45) $$
\noindent In order to make use of the heat-kernel technique, while the heat equation is of \textit{second order} in the differential operator, we need to re-formulate $\Gamma_{fermion-loop}$ so that the differential operator in the determinant is of second order. This can be accomplished by replacing in (45) the linear differential operator $i\gamma^{\mu}\hat{D}_{\mu}$ by its square $(i\gamma^{\mu}\hat{D}_{\mu})^{2}$ and multiplying an overall factor of $\frac{1}{2}$, namely,
$$\Gamma_{fermion-loop}=-i\frac{1}{2}ln Det[(i\gamma^{\mu}\hat{D}_{\mu})^{2}]. \eqno (46) $$
\noindent The square in the determinant can be expressed as \cite{Nieh}
$$-(g^{\mu\nu}\hat{D}_{\mu}\hat{D}_{\nu}-\frac{i}{2}\sigma^{\mu\nu}C^{\lambda}_{\mu\nu}\hat{D}_{\lambda}+Z), \eqno (47) $$
\noindent where
$$Z=\frac{1}{4}R+\frac{i}{8}\gamma_{5}\tilde{R}+\frac{i}{2}\sigma^{\mu\nu}(R_{\mu\nu}+i\hat{F}_{\mu\nu}), \eqno (48) $$
\noindent with
$$R=R^{\mu\nu}_{~~\mu\nu}, $$
$$\tilde{R}=\varepsilon^{\mu\nu\lambda\rho}R_{\mu\nu\lambda\rho}, $$
$$R_{\mu\nu}=R^{\lambda}_{~\mu\lambda\nu}.$$
\noindent We note that when torsion vanishes, $\tilde{R}=0$, $R_{\mu\nu}=R_{\nu\mu}$, and Z in (48) reduces to $\frac{1}{4}R$, a well recognized result for Riemannian spacetime.
The linear derivative term in (47), which is proportional to the torsion tensor, can be absorbed into the quadratic derivative term by redefining the covariant derivative
$$\hat{\tilde{D}}_{\mu}=\hat{D}_{\mu}-\frac{i}{4}\sigma^{\lambda\rho}C_{\lambda\rho\mu}. \eqno (49) $$
\noindent The fermion-loop effective action (46) thus becomes
$$\Gamma_{fermion-loop}=-i\frac{1}{2}ln Det[g^{\mu\nu}\hat{\tilde{D}}_{\mu}\hat{\tilde{D}}_{\nu}+X]. \eqno (50) $$
\noindent with
$$X=Z-D_{\mu}Q^{\mu}-Q_{\mu}Q^{\mu}, \eqno (51) $$
\noindent where
$$Q_{\mu}=-\frac{i}{4}\sigma^{\lambda\rho}C_{\lambda\rho\mu}. \eqno (52) $$
The divergent part of the fermion-loop effective action (48) is of the form \cite{Parker}
$$Div\Gamma_{fermion-loop}=\frac{1}{\epsilon}\displaystyle\int d^{4}x h tr [a_{2}](x), \eqno (53) $$
\noindent where $\epsilon=(4\pi)^{2}(n-4)$ and the corresponding kernel for $[a_{2}]$ is $g^{\mu\nu}\hat{\tilde{D}}_{\mu}\hat{\tilde{D}}_{\nu}+X$ in (50). For Riemann-Cartan spacetime and in the case of totally antisymmetric torsion tensor, the $[a_{2}]$ corresponding to the differential operator in (47) has been obtained by Yajima \cite{Yajima}. It is given by, as adopted with our metric,
$$\begin{array}{rl}[a_{2}]=\frac{1}{12}\tilde{W}^{\mu\nu}\tilde{W}_{\mu\nu}+\frac{1}{180}(R^{(o)\mu\nu\lambda\rho}{R^{(o)}_{~\mu\nu\lambda\rho}}-R^{(o)\mu\nu}R^{(o)}_{~\mu\nu})\\
-\frac{1}{6}\hat{\tilde{D}}_{\mu}\hat{\tilde{D}}^{\mu}(\frac{1}{5}R^{(o)}-X)+\frac{1}{2}(\frac{1}{6}R^{(o)}-X)^{2}, \end{array}\eqno (54) $$
\noindent where $R^{(o)}_{~\mu\nu\lambda\rho}$, etc. are the Riemannian curvature tensors, and $\tilde{W}_{\mu\nu}$ is defined \cite{Yajima} according to
$$[\hat{\tilde{D}}_{\mu},\hat{\tilde{D}}_{\nu}]\psi=(\tilde{W}_{\mu\nu}+C^{\lambda}_{~\mu\nu}\hat{\tilde{D}}_{\lambda})\psi.   \eqno (55)  $$
\noindent We remark that the disentanglement with the definition of $\tilde{W}_{\mu\nu}$ of the torsion term in (55) is crucial in assuring gauge symmetry in the final result. With the definition in (47), $\tilde{W}_{\mu\nu}$ is computed to be
$$\begin{array}{rl}\tilde{W}_{\mu\nu}=-\frac{i}{4}\sigma^{\alpha\beta}[R^{(o)}_{~\alpha\beta\mu\nu}+\frac{3}{2}(\bar{\nabla}_{\mu}C_{\alpha\beta\nu}-\bar{\nabla}_{\nu}C_{\alpha\beta\mu}\\
-C^{~\lambda}_{\alpha~\mu}C_{\beta\lambda\nu}+C^{~\lambda}_{\alpha~\nu}C_{\beta\lambda\mu})]+i{\hat F}_{\mu\nu}, \end{array}\eqno (56)  $$
\noindent where $\bar{\nabla}_{\mu}$ is the Riemannian covariant derivative. We point out one important aspect of (53) is that the torsion term in (53) is not involved in the definition of $\tilde {W}_{\mu\nu}$. This ensures the clean appearance of the \textit{gauge invariant} $\hat{F}_{\mu\nu}$ term in $\tilde{W}_{\mu\nu}$ without involvement of the torsion tensor.
\noindent Trace over the spinor and quark indices of the quark field $\psi$ of the product $\tilde{W}^{\mu\nu}\tilde{W}_{\mu\nu}$ term appearing in (54) is given by
$$\frac{1}{12}tr \tilde{W}^{\mu\nu}\tilde{W}_{\mu\nu}=-\frac{1}{8}\Sigma^{\alpha\beta\mu\nu}\Sigma_{\alpha\beta\mu\nu}-\frac{1}{3}g^{2}tr {\hat F}^{\mu\nu}{\hat F}_{\mu\nu}, \eqno (57) $$
\noindent where
$$\begin{array}{rl}\Sigma_{\alpha\beta\mu\nu}=R^{(o)}_{~\alpha\beta\mu\nu}+\frac{3}{2}(\bar{\nabla}_{\mu}C_{\alpha\beta\nu}-\bar{\nabla}_{\nu}C_{\alpha\beta\mu}\\
-C^{~\lambda}_{\alpha~\mu}C_{\beta\lambda\nu}+C^{~\lambda}_{\alpha~\nu}C_{\beta\lambda\mu}). \end{array}\eqno (58)  $$
As our main interest is in the renormalizabity of gauge theory in the Riemann-Cartan spacetime, we will concentrate on the gauge-field terms in (53). Explicit evaluation shows that these terms come from the first and last terms on the right-hand side of (54). The contribution from the first term is contained in (57). The contribution from the last term is in
$$tr \frac{1}{2}X^{2}=g^{2}tr{\hat F}^{\mu\nu}{\hat F}_{\mu\nu}+gravitational~terms.  $$
\noindent The renormalization counter term for the gluon field due to the fermion loop is the sum of the two contributions and is given by
$$\begin{array}{rl}Div\Gamma_{fermion-loop}=\frac{1}{\epsilon}\displaystyle\int d^{4}x h (\frac{2}{3}g^{2}tr \hat{F^{\mu\nu}}\hat{F_{\mu\nu}}\\
+gravitational~terms).\end{array} \eqno (59) $$

\section{Gluon and Ghost Loops}

The relevant terms for the gluon loop in $W^{(2)}$ are
$$\begin{array}{rl} W_{gluon-loop}=\displaystyle\int d^{4}x \\h\{-\frac{1}{4}\tilde{F}^{\mu\nu}\tilde{F}_{\mu\nu}-2iA_{\mu}[\hat{F}^{\mu\nu},A_{\nu}]
-\frac{1}{2}(\hat{\nabla}_{\mu}A^{\mu})^{2}\}. \end{array}\eqno (60) $$
\noindent while that for the ghost loop is
$$W_{ghost-loop}=\displaystyle\int d^{4}x h\bar{\zeta}(-\hat{\nabla}_{\mu}\hat{\nabla}^{\mu})\zeta.  \eqno (61) $$
\noindent Up to a total derivative, $W_{gluon-loop}$ can be expressed in the following form,
$$\begin{array}{rl}\displaystyle\int d^{4}x \frac{1}{2}hA^{\mu}(g_{\mu\nu}\hat{\nabla}_{\lambda}\hat{\nabla}^{\lambda}-\hat{\nabla}_{\nu}\hat{\nabla}_{\mu}+\hat{\nabla}_{\mu}\hat{\nabla}_{\nu}\\
+2C_{\mu\lambda\nu}\hat{\nabla}^{\lambda}+\hat{\nabla}^{\lambda}C_{\mu\lambda\nu}-C_{\mu\lambda\rho}C_{\nu}^{~\lambda\rho}+g\hat{F}_{\mu\nu})A^{\nu}. \end{array}
\eqno (62) $$
\noindent where $\hat{\nabla}_{\mu}$ is defined as in
$$\hat{\nabla}_{\mu}A_{\nu}=\nabla_{\mu}A_{\nu}-i[\hat{A_{\mu}},A_{\nu}]. \eqno (63)$$
\noindent In (62), the $A\hat{F}A$ product term is understood to be the product $f^{\underline{abc}}A^{\underline{a}}\hat{F}^{\underline{b}}A^{\underline{c}}$. On account of
$$[\hat{\nabla}_{\mu},\hat{\nabla}_{\nu}]A^{\nu}=R_{\nu\mu}A^{\nu}+C^{\rho}_{~\mu\nu}\nabla_{\rho}A^{\nu}-ig[\hat{F}_{\mu\nu},A^{\nu}], \eqno (64) $$
\noindent we can express (62) in the form
$$\begin{array}{rl}\frac{1}{2}hA^{\mu}[g_{\mu\nu}\hat{\nabla}_{\lambda}\hat{\nabla}^{\lambda}+C_{\mu\lambda\nu}\hat{\nabla}^{\lambda}
+\hat{\nabla}^{\lambda}C_{\mu\lambda\nu}-C_{\mu\lambda\rho}C_{\nu}^{~\lambda\rho}\\+R_{\nu\mu}+2g\hat{F}_{\mu\nu}]A^{\nu}. \end{array}\eqno (65), $$
\noindent The term linear in derivative in (64), which is brought about by torsion, is to be absorbed into the quadratic derivative term by defining a modified connection
$$\hat{\Gamma}^{'\lambda}_{~~\mu\nu}=\hat{\Gamma}^{\lambda}_{~\mu\nu}+\frac{1}{2}C^{\lambda}_{~\mu\nu}, \eqno (66) $$
\noindent with the corresponding covariant derivative expressed as $\hat{\nabla}{'}_{\mu}$.  The gluon-loop action (62) can be written as
$$W_{gluon-loop}=\displaystyle\int d^{4}x\frac{1}{2}hA^{\mu}(g_{\mu\nu}\hat{\nabla}^{'}_{~\lambda}\hat{\nabla}^{'\lambda}+X_{\mu\nu})A^{\nu}, \eqno (67) $$
\noindent where
$$X^{\underline{ac}}_{\mu\nu}=R_{\nu\mu}\delta^{\underline{ac}}+2gf^{\underline{abc}}\hat{F}^{\underline{b}}_{\mu\nu}. \eqno (68) $$
\noindent The gluon-loop effective action $\Gamma_{gluon-loop}$ is then given by
$$\Gamma_{gluon-loop}=i\frac{1}{2}ln det(g_{\mu\nu}\hat{\nabla}^{'}_{\lambda}\hat{\nabla}^{'\lambda}+X^{'}). \eqno (69) $$
\noindent Its divergent pole term is
$$Div\Gamma_{gluon-loop}=-\frac{1}{\epsilon}\displaystyle\int d^{4}x htr [a^{'}_{2}](x), \eqno (70) $$
\noindent where $[a^{'}_{2}]$ is the asymptotic expansion coefficient corresponding to the kernel appearing in (67), with a structure similar to that for the fermion case, namely, as in (54). Again, we will concentrate on the corresponding first and last terms in (54) that give rise to gauge-field terms.
\noindent The corresponding $W^{'}_{~\mu\nu}$ can be obtained from calculating $[\hat{\nabla}^{'}_{~\mu},\hat{\nabla}^{'}_{~\nu}]A_{\lambda}$, which can be neatly expressed as
$$[\hat{\nabla}^{'}_{~\mu},\hat{\nabla}^{'}_{~\nu}]A_{\lambda}=R^{'\rho}_{~~\lambda\mu\nu}A_{\rho}-i[\hat{F_{\mu\nu}},A_{\lambda}]+C^{'\rho}_{~~\mu\nu}\hat{\nabla}^{'}_{~\rho}A_{\lambda}, \eqno (71) $$
\noindent where $R^{'\rho}_{~~\lambda\mu\nu}$ is the curvature tensor formed with $\Gamma^{'\lambda}_{~~\mu\nu}$ as the connection. From (71) we obtain
$$(W^{'}_{~\mu\nu})^{\rho\underline{ac}}_{~\lambda}=R{'\rho}_{~~\lambda\mu\nu}\delta^{\underline{ac}}+gf^{\underline{abc}}{\hat F}^{\underline{b}}_{\mu\nu}\delta^{\rho}_{~\lambda}.  \eqno (72)  $$
\noindent We then obtain contribution from the first term,
$$\begin{array}{rl}tr(\frac{1}{12}W^{'}_{~\mu\nu}W^{'\mu\nu})=-\frac{1}{3}g^{2}f^{\underline{abc}}f^{\underline{cda}}\hat{F}^{\underline{b}\mu\nu}\hat{F}^{\underline{d}}_{\mu\nu}\\
+gravitational~terms. \end{array}\eqno (73)$$
\noindent Define $C_{2}(G)$ by
$$f^{\underline{abc}}T^{\underline{a}}T^{\underline{c}}=\frac{i}{2}C_{2}(G)T^{\underline{b}}. \eqno (74) $$
\noindent We then have
$$\begin{array}{rl}tr(\frac{1}{12}W^{'}_{~\mu\nu}W^{'\mu\nu})=-\frac{1}{3}g^{2}\frac{C_{2}(G)}{C_{2}(R)}tr \hat{F}^{\mu\nu}\hat{F}_{\mu\nu}\\
+gravitational~terms, \end{array}\eqno (75)$$

\noindent where $C_{2}(R)$ is the normalization factor given by (20). The contribution from the last term can be similarly calculated and is given by
$$tr\frac{1}{2}X_{\mu\nu}X^{\nu\mu}=2g^{2}\frac{C_{2}(G)}{C_{2}(R)}\hat{F}^{\mu\nu}\hat{F}_{\mu\nu}+gravitational~terms. \eqno (76) $$
\noindent The divergent pole term of the gluon loop is due to the sum of the contributions in (75) and (76) and given by
$$\begin{array}{rl}Div\Gamma_{gluon-loop}=-\frac{1}{\epsilon}\displaystyle\int d^{4}x h(\frac{5}{3}g^{2}\frac{C_{2}(G)}{C_{2}(R)}tr \hat{F}^{\mu\nu}\hat{F}_{\mu\nu}\\
+gravitational~terms). \end{array}\eqno (77)  $$
The effective action due to the loop of the \textit{complex} Grassmann ghost field is given by
$$\Gamma_{ghost-loop}=-2iln det(-\hat{\nabla}_{\mu}\hat{\nabla}^{\mu}). \eqno (78) $$
\noindent Its divergent part can be similarly calculated, taking into account that the ghost field is a complex Grassmann scalar and there is no spinor index to sum over, is given by
$$\begin{array}{rl}Div\Gamma_{ghost-loop}=\frac{1}{\epsilon}\displaystyle\int d^{4}x h(-\frac{1}{6}g^{2}\frac{C_{2}(G)}{C_{2}(R)}tr \hat{F}^{\mu\nu}\hat{F}_{\mu\nu}\\
+gravitational~terms). \end{array}\eqno (79)$$
The final result of the divergent parts due to the quark-, gluon-, and ghost-loops is the sum of (59), (77) and (79)
$$\begin{array}{rl}Div\Gamma_{loops}=\frac{1}{\epsilon}\displaystyle\int d^{4}x hg^{2}tr \frac{4C_{2}(R)-11C_{2}(G)}{6C_{2}(R)}tr \hat{F}^{\mu\nu}\hat{F}_{\mu\nu}\\
+gravitational terms. \end{array}\eqno (80) $$

\noindent We recall that we have for convenience omitted the normalization factor of $\frac{1}{C_{2}(R)}$ for the $F^{\mu\nu}F_{\mu\nu}$ term in the original Lagrangian in (17). Thus, when we consider renormalization constants, this normarlization factor should be similarly omitted, namely, by dropping $C_{2}(R)$ in the denominator in the above equation. The resulting result for the gauge-field term, we note, is compatible with earlier results \cite{Toms,Omote} for the Riemannian spacetime. In our specific case of quantum chromodynamics without additional flavor, $C_{2}(G)=3$ and $C_{2}(F)=\frac{1}{2}$.

\section{Mixed Loops}

These are the gluon-quark vertex and quark self-energy loops, which contain both internal quark and gluon lines. We will use 't Hooft's algorithm \cite{'t Hooft, Omote} to find the renormalization counter terms. Following 't Hooft's procedure, we make the substitutions in the action $W^{(2)}$ in (41)

$$\bar{\psi}\rightarrow\bar{\psi},  $$
$$\psi\rightarrow\gamma^{\mu}\acute{D}_{\mu}\xi,  $$

\noindent where $\acute{D}_{\mu}=\partial_{\mu}-\frac{i}{4}\sigma_{ab}\omega_{~~\mu}^{ab}$, namely, the covariant derivative $D_{\mu}$ without the gauge-potential term. The fermion part of the Lagrangian in (41) becomes

$$\begin{array}{rl}h[-g^{\mu\nu}\bar{\psi}\acute{D}_{\mu}\acute{D}_{\nu}\xi+
\frac{i}{2}\sigma^{\mu\nu}C^{\lambda}_{~\mu\nu}\acute{D}_{\lambda}\xi-\bar{\psi}\acute{Z}\xi\\
+\bar{\psi}i\gamma^{\mu}ig\hat{A}_{\mu}i\gamma^{\nu}\acute{D}_{\nu}\xi\\
+\bar{\psi}i\gamma^{\mu}igA_{\mu}\eta+\bar{\eta}i\gamma^{\mu}igA_{\mu}i\gamma^{\nu}\acute{D}_{\nu}\xi]. \end{array}\eqno (81) $$

\noindent Applying 't Hooft's algorithm to this fermion Lagrangian and the gluon part of the Lagrangian as given in (67), we have, in 't Hooft's notation,

$$(\alpha)^{\underline{a}}_{\lambda}=i\gamma_{\lambda}igT^{\underline{a}}\eta,  $$
$$\bar{\beta}^{\lambda\underline{b}}=\bar{\eta}i\gamma^{\lambda}iigT^{\underline{b}},  $$
$$(N^{\mu})^{\rho\underline{ac}}_{~\lambda}=g^{\mu\nu}(\hat{\Gamma^{'\rho}_{~~\lambda\nu}}\delta^{\underline{ac}}+gf^{\underline{abc}}\hat{A}^{\underline{b}}_{\nu}g^{\rho}_{~\lambda}),  $$

\noindent where $\gamma^{\mu}=e_{\underline{a}}^{~\mu}\gamma^{\underline{a}}$. According to the algorithm, the renormalization counter terms due to the mixed loops are the sum of the following four terms:

$$\frac{2}{\epsilon}\frac{1}{2}\bar{\beta}\gamma^{\mu}\partial_{\mu}\alpha, \eqno (82)  $$
$$\frac{2}{\epsilon}\frac{1}{2}[-\bar{\beta}\gamma^{\mu}(-\frac{i}{4}\sigma_{ab}\omega_{~~\mu}^{ab}-\frac{i}{4}\sigma^{\lambda\rho}C_{\lambda\rho\mu})]\alpha, \eqno (83)  $$
$$\frac{2}{\epsilon}\frac{1}{2}N_{\mu}\frac{1}{2}\bar{\beta}\gamma^{\mu}\alpha,  \eqno (84)  $$
$$\frac{2}{\epsilon}\frac{1}{2}\bar{\beta}\gamma^{\mu}\frac{1}{2}i\gamma^{\nu}\hat{A}_{\nu}\gamma^{\mu}i\alpha.  \eqno (85)  $$

\noindent With the help of the relations

$$\gamma_{a}\omega^{ba}_{~~\mu}e_{b}^{~\nu}=\frac{i}{4}\omega_{bc\mu}[\sigma^{bc},\gamma^{\nu}], \eqno (86)  $$
$$\partial_{\mu}\gamma^{\nu}=[\frac{i}{4}\sigma_{ab}\omega_{~~\mu}^{ab},\gamma^{\nu}]-\Gamma^{\nu}_{~\lambda\mu}\gamma^{\lambda},  \eqno (87)  $$
$$\partial_{\mu}\gamma_{\nu}=[\frac{i}{4}\sigma_{ab}\omega_{~~\mu}^{ab},\gamma_{\nu}]+\Gamma^{\lambda}_{~\nu\mu}\gamma_{\lambda},  \eqno (88)  $$

\noindent the sum of the four terms is given by

$$\begin{array}{rl}\frac{2}{\epsilon}[\bar{\eta}g^{2}(T^{\underline{a}}T^{\underline{a}})i\gamma^{\mu}(\partial_{\mu}-\frac{i}{4}\sigma_{ab}\omega_{~~\mu}^{ab})\eta\\
-\bar{\eta}g^{3}(f^{\underline{abc}}T^{\underline{a}}T^{\underline{c}})\hat{A}^{\underline{b}}_{\nu}i\gamma^{\nu}\eta
+\bar{\eta}g^{3}(T^{\underline{a}}T^{\underline{b}}T^{\underline{a}})i\hat{A}^{\underline{b}}_{\mu}i\gamma^{\mu}\eta].\end{array}\eqno (89)  $$

\noindent In addition to $C_{2}(R)$ defined by (20) and $C_{2}(G)$ defined by (74), we further define $C_{2}(F)$ by \cite{Ryder}

$$T^{\underline{a}}T^{\underline{a}}=C_{2}(F)I. \eqno(90) $$

\noindent We note that $C_{2}(F)$ and $C_{2}(R)$ are related. In our specific case here, $C_{2}(F)=\frac{8}{3}C_{2}(R)=\frac{4}{3}$. It can be easily shown that

$$f^{\underline{abc}}T^{\underline{a}}T^{\underline{c}}=\frac{i}{2}C_{2}(G)T^{\underline{b}}, $$

$$T^{\underline{a}}T^{\underline{b}}T^{\underline{a}}=-\frac{1}{2}C_{2}(G)T^{\underline{b}}+C^{2}(F)T^{\underline{b}}.  $$

\noindent The sum (89) becomes,

$$\frac{2}{\epsilon}C_{2}(F)g^{2}\bar{\eta}i\gamma^{\mu}(\partial_{\mu}-\frac{i}{4}\sigma_{ab}\omega_{~~\mu}^{ab}+ig\hat{A}_{\mu})\eta. \eqno (91) $$

\noindent This is the final result for the renormalization counter terms due to the gluon-quark vertex and quark self-energy loops. It is also compatible with the earlier result \cite{Toms, Omote} for the Riemannian spacetime.
\\
\\
\section{Conclusions}

We have in this note deliberated on the compatibility of torsion with gauge symmetry in a realistic interacting gauge theory, namely, quantum chromodynamics of gluons interacting with quarks in Riemann-Cartan spacetime. We have demonstrated that the system of classical field equations are consistent with the choice of the gauge invariant definition of $F_{\mu\nu}$, which is the flat-space expression, while inconsistent with the choice of the gauge non-invariant version, which is the one with covariant derivatives replacing the ordinary derivatives in the flat-space expression. To further substantiate the choice of the gauge invariant version of $F_{\mu\nu}$, we have investigated the quantum renormalizability at on-loop level, to make sure that torsion does not somehow get entangled with gauge symmetry at a level beyond the classical. The heat-kernel technique being an essential method in our treatment, we restrict ourselves to the special case of totally antisymmetric torsion, as the general case is much more complicated and there is lack of reliable study on the corresponding heat kernel. We would like to note that the results of Yajima and collaborators \cite{Yajima} on the heat kernel in the presence of torsion are essential for our results.

With regard to the renormalization counter terms for the gluon field and the quark field, our one-loop results are contained in (80) and (91). It is seen that the counter terms are in the same gauge invariant forms as the original terms in the Lagrangian. Except for the gravitational counter terms, which we have omitted, the pattern of counter terms for the gluon and quark fields in the present case of Riemann-Cartan spacetime is exactly the same as in the previously studied case of Riemannian spacetime \cite{Toms, Omote}. We will hence not repeat here defining the renormalized constants. The conclusion, of course, is that the theory, with the choice of the gauge invariant version of $F_{\mu\nu}$, is renormalizable and gauge symmetry preserved at the one-loop level.

\vspace{2mm}
\begin{acknowledgements}

The author wishes to thank Professor Friedrich Hehl for communications regarding gauge symmetry and his critical comments. He would also like to thank Professor S. Yajima for clarification on the results published by his group.

\end{acknowledgements}


\begin{thebibliography}{99}
\bibitem{Higgs}  We remark that in the standard-model gauge theories, the fermion mass is due to the Brout-Englert-Higgs mechanism of spontaneous symmetry breaking.
\end{thebibliography}
\end{document}